\documentclass[a4paper,10pt, twoside]{article}
\title{Operads and the Hopf algebras of renormalisation}
\author{Pepijn van der Laan}

\usepackage[english]{babel}
\usepackage{amsmath, amssymb,theorem}
\usepackage[all]{xy}
\usepackage{epsfig,color}
\UseComputerModernTips
\newcommand{\note}{\textbf}
\theoremheaderfont{\normalfont\bfseries}
\theorembodyfont{\normalfont\itshape}
\theoremstyle{change}
\newtheorem{Tm}{\normalfont\scshape Theorem}[section]
\newtheorem{Pp}[Tm]{\normalfont\scshape Proposition}
\newtheorem{Lm}[Tm]{\normalfont\scshape Lemma}
\newtheorem{Cr}[Tm]{\normalfont\scshape Corollary}

\theorembodyfont{\rmfamily}

\newtheorem{Df}[Tm]{\normalfont\scshape Definition}
\newtheorem{Ex}[Tm]{\normalfont\scshape Example}
\newtheorem{Rm}[Tm]{\normalfont\scshape Remark}

\newenvironment{Pf}{{\par\scshape Proof
}}{\hspace*{\fill}{\scshape QED}\\ \vspace*{0.5cm}\par}
\makeatletter
\renewcommand\subsection{\@startsection{subsection}{2}{\z@}%
                                     {-3.3ex\@plus -1ex \@minus -.2ex}%
                                     {2ex \@plus .2ex}%
                                     {\centering\normalfont\bfseries}}
\renewcommand{\@makecaption}[2]{%
\vspace{10pt}\hspace{.1\linewidth}
\parbox[l]{.8\linewidth}{\footnotesize{\scshape #1:} #2}%
\par
}
\makeatother
\newcommand{\NN}{\mathbb{N}}
\newcommand{\ZZ}{\mathbb{Z}}

\renewcommand{\phi}{\varphi}

\newcommand{\eps}{\varepsilon}

\newcommand{\cobar}{\smash{\raise.5\baselineskip\hbox{\begin{turn}{180}$B$\end{turn}}}}
\newcommand{\pd}{\partial}

\newcommand{\End}{\mathrm{End}}
\newcommand{\Aut}{\mathrm{Aut}}
\newcommand{\id}{\mathrm{id}}

\newcommand{\Hom}{\mathrm{Hom}}
\renewcommand{\vert}{\textbf{v}}
\newcommand{\edge}{\textbf{e}}
\newcommand{\leg}{\textbf{l}}

\newcommand{\hoa}{Hopf algebra}

\begin{document}
\maketitle
\abstract{Functors from (co)operads to
bialgebras relate Hopf algebras that occur in renormalisation
to operads, which simplifies the proof of the Hopf algebra axioms, and
induces a characterisation of the corresponding group of characters and Lie
algebra of primitives of the dual in terms of the operad.  
In addition, it is shown that the Wick rotation formula leads to
canonical algebras for one of these operads. 
}

\section{Introduction}
Kreimer \cite{Krei:Hoa} observes that a governing principle of
renormalisation is given by the antipode of a Hopf algebra. More
Hopf algebras related to renormalisation have been defined since
then. Apart from checking the Hopf algebra axioms, one is interested
in the group of characters and the Lie algebra of primitive 
elements of the dual of these Hopf algebras. This paper shows that these
coproducts often decompose into more elementary operations which make
checking of the Hopf algebra axioms less cumbersome. Moreover, one
obtains a description of the group of characters and the Lie algebra of
primitive elements of the dual. 

Section \ref{Sec:constructions} defines functors $C\longmapsto H_C$
and $C\longmapsto\bar H_C$ from 
1-reduced cooperads to connected complete Hopf algebras. This section
shows that for $C$ of finite type, the character groups
$\Hom_{\text{Alg}}(H_C,k)$ and $\Hom_{\text{Alg}}(\bar H_C,k)$ of
$H_C$ and $\bar H_C$ are given as explicit functors from operads to
groups applied to the dual operad of $C$. The Lie algebra of primitive
elements of the dual Hopf algebra is given by well known functors from
operads to Lie algebras applied to the the dual operad of $C$. 

Section \ref{Sec:firstexamples} explores some examples. 
The bitensor algebra of a bialgebra and its Pinter Hopf algebra are
obtained using the constructions of section \ref{Sec:constructions}
(cf. Van der Laan-Moerdijk \cite{IekePep:Hopf}). The
Hopf algebra of higher order differentials of the line, and its non-commutative
version (e.g. Brouder-Frabetti \cite{BrouFrab:Noncom}) are obtained
from the construction applied to the operad of commutative
algebras.

Section \ref{Sec:operadofgraphs} is devoted to one specific
example. This section
constructs an operad of graphs and shows that the Connes-Kreimer
Hopf algebra of graphs \cite{ConKrei:RieHilI} is the Hopf algebra
constructed from the suboperad of one particle irreducible graphs.


\subsection{Acknowledgements}

I am grateful to Alessandra Frabetti for the many discussions that
were of great help, and for her enthousiasm.
I thank Ieke Moerdijk, and  Muriel Livernet and Christian
Brouder for the enjoyable and helpful discussions. The research is
part of my Ph.D.-thesis \cite{Pep:thesis}, and was partly
supported by Marie Curie Training Site Fellowships HPMT-CT-2001-00367
(Universit\'e Paris Nord XIII) and HPMT-2000-00075 (Centre de Recerca
Matem\`atica, Barcelona).

\section{Preliminaries}

Throughout this article we work in the category of vector spaces over a
field $k$ of characteristic 0. 

By $S_n$ we denote the symmetric group on $n$ letters, and by 
$kS_n$ its group algebra which is the vector space spanned by the set
$S_n$ whose multiplication is the linear extension of multiplication
in $S_n$. If $S_n$ acts on a vector space $V$, the coinvariants of the
group action are denoted $V_{S_n}$ and the invariants by $V^{S_n}$.

For a vector space $V$ we will denote by $TV =
\bigoplus_{n\geq 0} V^{\otimes n}$ the unital tensor algebra on $V$
with the concatenation product, and by $SV = \bigoplus_{n\geq 0}
(V^{\otimes n})_{S_n}$ the unital symmetric algebra on $V$  which is
the quotient of $T_V$ by its commutator ideal $[TV,TV]$. Both $SV$ and
$TV$ are graded algebras with respect to $n$. If $W = \bigoplus_n W_n$
is a graded vector space, denote by $W^* =
\bigoplus_nW_{n}^*$ its graded dual.

\subsection{Operads}

A non-symmetric operad is a sequence
$\{P(n)\}_{n\geq 1}$ of vector spaces together with composition maps  
\[
\gamma:P(n)\otimes P(m_1)\otimes\ldots\otimes P(m_n) \longrightarrow
P(m_1+\ldots+ m_n),
\]
and an identity element $\id\in P(1)$. These structures satisfy the
following axioms:
\begin{enumerate}
\item
The composition satisfies the associativity relation: If we write
\[
p(q_1,\ldots,q_n) := \gamma(p,q_1,\ldots,q_n)
\] for $p\in P(n)$, and $q_i\in P(m_i)$
 for $i=1,\ldots, n$, then for any such elements
$p$ and $q_i$,and any $r_j\in P(k_j)$ for
$j=1,\ldots,m_1+\ldots+m_n$ and for $k_j\geq 0$ the following holds:
\[
\begin{split}
&(p(q_1,\ldots,q_n))(r_1,\ldots, r_{m_1+\ldots + m_n}) =\\ 
&\ p(q_1(r_1,\ldots,r_{m_1}),\ldots, q_n(r_{m_1+\ldots+
  m_{n-1}+1},\ldots, r_{m_1+\ldots+ m_{n}})).
\end{split}
\]
\item
The identity element $\id\in P(1)$ acts as a left and right identity:
If $p\in P(n)$, then 
\[
\id(p) = p = p(\id,\ldots,\id).
\]
\end{enumerate}
There is a natural grading on the
total space $\bigoplus_nP(n)$, defined by 
\[
(\bigoplus_nP(n))^m =P(m+1).
\] 

A collection $P$ is a sequence of vector spaces $\{P(n)\}_{n\geq 1}$
such that each $P(n)$ has a right $S_n$-module structure. An
(symmetric) operad is a collection $P$ together with an
non-symmetric operad structure on the sequence of vector spaces, and
\begin{itemize}
\item[(iii)]
composition is
equivariant with respect to the $S_n$-actions is the sense that
\[
\begin{split}
p\sigma(q_1,\ldots,q_n)  &= p(q_{\sigma(1)},\ldots,q_{\sigma(n)})\hat\sigma\\
p(q_1\sigma_1,\ldots,q_n\sigma_n) &=
p(q_1,\ldots,q_n)(\sigma_1\times\ldots\times \sigma_n)
\end{split}
\]
for any $p\in P(n),\ \sigma\in S_n$ and $q_i\in P(m_i),\ \sigma_i\in
S_{m_i}$ for $i = 1\ldots n$, where $\hat\sigma$ is the permutation on
$m_1+\ldots+m_n$ elements that permutes consecutive blocks of lengths
$m_1, m_2,\ldots, m_n$ according to $\sigma\in S_n$.
\end{itemize}
Dually (in the sense of inverting direction of arrows in the defining
diagrams), one defines (non-symmetric) cooperads. In particular, a
non-symmetric cooperad $C$ is a sequence of vetor spaces
$\{C(n)\}_{n\geq 1}$, together with cocopmosition maps
\[
\gamma^*:C(n)\longrightarrow \bigoplus_k\bigoplus_{n_1+\ldots+n_k = n} 
C(k) \otimes C(n_1)\otimes \ldots \otimes C(n_k),
\]
and a coidentity $\id^*:C(1)\longrightarrow k$ that satisfy the dual
relations. A Cooperad is a non-symmetric cooperad together with a collection
structure on $C$ that makes $\gamma^*$ an equivariant map. 

For more extensive background on (co)operads read Getzler-Jones
\cite{GetzJon:Opd}, and Markl-Shnider-Stasheff 
\cite{MarShniSta:Opd}.

\subsection{Graphs}

A graph $\eta$ consist of sets $\vert(\eta)$ of vertices, a set
$\edge(\eta)$ of internal edges, and a set $\leg(\eta)$ of external
edges or legs; together with a map that assigns to each edge a pair of
(not necessary distinct) vertices and a map that assigns to each leg a
vertex. To draw a graph, draw a dot for each vertex $v$, and for each
edge $e$ draw aline between the two vertices assigned to it, and for
each leg draw a line that in one end ends in the vertex assigned to
it. If $v\in \vert(\eta)$, denote by $\leg(v)\subset
\edge(v)\cup\leg(\eta)$ the set of legs and edges attached to
$v$. Call these the legs of $v$. A morphism of graphs consists of
morphisms of vertices, edges, and legs compatible with the structure
maps. 

A connected graph $t$ is a tree if $|\vert(t)| = |\edge(t)|+1$. A
rooted tree is a tree together with a basepoint $r\in\leg(t)$, the
root. 

\section{Constructions on cooperads}
\label{Sec:constructions}

\subsection{Bialgebras}\label{subsec:Hopfalgebras}

\begin{Df}
Let $C$ be a non-symmetric cooperad. We denote $B_C =
T(\bigoplus_nC(n))$ the tensor algebra on the total space of $C$. Use
the cocomposition 
\[
\gamma^*:C(n) \longrightarrow \bigoplus_{k,n_1+\ldots+n_k = n}C(k)
\otimes (C(n_1)\otimes\ldots\otimes C(n_k))
\]
and the natural inclusions 
\[
i_1:C(k)\longrightarrow T(\bigoplus_mC(m)), \quad\text{and}\quad
i_2:C(n_1)\otimes\ldots\otimes C(n_k)\longrightarrow T(\bigoplus_mC(m))
\] 
to define  a map $\Delta:B_C\longrightarrow B_C\otimes B_C$ on
generators as $\Delta = (i_1\otimes i_2)\circ\gamma^*$.
Extend $\Delta$ as an algebra morphism. Define the algebra morphism
$\eps:B_C\longrightarrow k$ as the map $\eps$ which vanishes on
generators of degree $\neq 0$, and satisfies $\eps|_{C(1)} = \eps_C$. 
\end{Df} 
\begin{Lm}\label{Lm:bialgebra}
Let $C$ be a non-symmetric cooperad. 
\begin{enumerate}
\item
Comultiplication $\Delta$ and counit $\eps$ as defined above make
$C\longmapsto B_C$ a functor from non-symmetric cooperads to graded
bialgebras.
\item
If $C$ is a cooperad the bialgebra structure of $B_C$ descends to
$\bar B_C = S(\bigoplus_nC(n)^{S_n})$ (the symmetric algebra on
the total space of invariants of $C$), and consequently $C\longmapsto
\bar B_C$ defines a functor from cooperads to commutative graded
bialgebras. 
\end{enumerate}
\end{Lm} 
A detailed proof can be found in Van der Laan \cite{Pep:thesis}.

\subsection{Hopf algebras}

A (co)operad $P$ is called \note{1-reduced}, or \note{1-connected} if
$P(0)= 0$, and $P(1) = k$. Let $C$ be a 1-reduced non-symmetric cooperad. The space
$\bigoplus_nC(n)$ has a base-point given by the inverse of the counit
$\varepsilon:C(1)\longrightarrow k$ . In the sequel we will 
use pointed tensor algebra and the pointed symmetric algebra 
\[
\begin{split}
T_*(\bigoplus_n C(n)) &=
T(\bigoplus_n C(n))/(\varepsilon^{-1}(1)-\mathbf{1})\\
S_*(\bigoplus_n C(n)) &=
S(\bigoplus_n C(n))/(\varepsilon^{-1}(1)-\mathbf{1}),
\end{split}
\]
where the unit in $T$ and $S$ is denoted by $\mathbf{1}$. In other
words, $T_*$ is the left adjoint to the forgetful functor
$(A,\mu,u)\longmapsto (A,u)$ from unital associative algebras to
vector spaces with a non-zero base-point, and $S_*$ the left adjoint
to the forgetful functor from unital commutative algebras to vector
spaces with non-zero base-point. 
\begin{Df}
Let $C$ be a 1-reduced non-symmetric cooperad.
Let 
\[H_C = T_*(\bigoplus_nC(n))\]
the pointed tensor algebra on the
total space of $C$, with respect
to the base-point given by the inclusion of $C(1)=k$. The coalgebra
structure on $B_C$ induces maps $\Delta:H_C\longrightarrow H_C\otimes
H_C$, and $\eps:H_C\longrightarrow k$. If $C$ is a 1-reduced cooperad, 
denote \[\bar H_C = S_*(\bigoplus_nC(n)^{S_n})\] the pointed symmetric
algebra on the total space of invariants of $C$. Lemma
\ref{Lm:bialgebra} now can be adapted to a pointed version.
\end{Df}
\begin{Tm}\label{Lm:Hopfalgebra}
Let $C$ be a 1-reduced non-symmetric cooperad. 
\begin{enumerate}
\item
The application $C\longmapsto H_C$ defines a functor from non-symmetric
cooperads to graded connected Hopf algebras. 
\item
If $C$ is a 1-reduced cooperad the application $C\longmapsto \bar H_C$
defines a functor from cooperads to commutative graded connected Hopf
algebras.
\end{enumerate}
\end{Tm}
\begin{Pf}
There is a natural surjection of algebras \[T(\bigoplus_nC(n))\longrightarrow
T_*(\bigoplus_{n}C(n)).\]
We define a coproduct on $T_*(\bigoplus_nC(n))$ by the formula of the
coproduct in Lemma \ref{Lm:bialgebra}.  
Since $C$ is coaugmented and $\eps$ is the counit, the cocomposition
$\gamma^*$ respects the base point given by
$\eps^{-1}:k\longrightarrow C(1)$, and satisfies
$(\eps\otimes \id)\circ \gamma^* = \id = (\id\otimes \eps)\circ
\gamma^*$. This implies the bialgebra structure is well defined on
the pointed tensor algebra. Functoriality is again trivial. To prove
\textit{(i)} it remains to check that $H_C$ is in fact a graded
connected Hopf algebra.

A bialgebra $A$ is called \note{connected} if it is $\ZZ$-graded,
concentrated in non-negative degree, and satisfies $A^0=k\cdot 1$. The
\note{augmentation ideal} of a connected bialgebra $A$ is the ideal
$\bigoplus_{n\geq 1} A^n$.
Since the degree 0 part of $H_C$ is $C(1)=k$, the bialgebra $H_C$ is
connected. 
It is well known (cf. Milnor and Moore \cite{MilnorMoore:Hoa}) that  
any graded connected bialgebra admits an antipode and is thus a Hopf
algebra. 

To pass to the symmetric version \textit{(ii)}, argue as in the proof
of Lemma \ref{Lm:bialgebra}, and observe that $\bar H_C$ is also
connected. 
\end{Pf}

\begin{Rm}
Note that the construction of $H_C$ and $\bar H_C$ defines a graded
bialgebra for any coaugmented cooperad $C$, but that $C$ needs to be
1-reduced in order to get a graded connected Hopf algebra.

Let $P$ be a (1-reduced) (non-symmetric) operad of finite type. Then
the linear dual collection $P^*$ is a (1-reduced) (non-symmetric)
cooperad. Regarding the bialgebras and Hopf algebras associated to the
cooperad $P^*$ I will use notation 
\[
B_P := B_{P^*}, \quad \bar B_P := \bar B_{P^*},\quad  
H_P := H_{P^*},\quad \text{and} \quad 
\bar H_P := \bar H_{P^*}.
\]
\end{Rm}


\subsection{Groups}

\begin{Df}
Let $P$ be a non-symmetric operad. Let
\[
G_P = \left\{\sum_{n=1}^\infty p_n\qquad \text{s.t.} \quad p_n\in
P(n)\text{ and }p_1 = \id \right\} \subset \hat{\bigoplus}_{n\geq 1}P(n),
\]
where $\hat{\bigoplus}$ denotes the completed sum.  Define a
multiplication $\circ$ on this set by 
\[
\left(\sum_np_n\right)\circ \left(\sum_mq_m\right) =
\sum_{n,m_1,\ldots,m_n} \gamma(p_n,q_{m_1},\ldots,q_{m_n}). 
\]
If $P$ is an operad this multiplication defines a
multiplication on
\[
\bar G_P = \left\{\sum_{n=1}^\infty p_n\qquad \text{s.t.} \quad p_n\in
P(n)_{S_n} \text{ and }p_1 = \id\right\} \subset
\hat\bigoplus_{n}(P(n)_{S_n}),  
\]
since the composition $\gamma$ is equivariant with respect to the
$S_n$-actions.
\end{Df}
\begin{Lm}
Let $P$ be a non-symmetric operad.
\begin{enumerate}
\item
The set $G_P$ is a group with respect to the multiplication $\circ$
the unit element $\id\in G_P$. The application $P\longrightarrow G_P$
defines a functor from non-$\sigma$ operads to groups.
\item 
If $P$ is an operad, the quotient $\bar G_P$ of $G_P$ is a group. The
application $P\longmapsto \bar G_P$ defines a functor from operads to
groups
\end{enumerate}
\end{Lm}
\begin{Pf}
Associativity is clear from the associativity of $\gamma$. That
$\id\in P(1)$ is the identity with respect to composition is also
obvious. To show the existence of an inverse, use a recursive
construction similar to the proof of the Formal Inverse Function Theorem.
\end{Pf}
\begin{Rm}
A group closely related to $G_P$ and $\bar G_P$ was defined
independently by Chapoton \cite{Chap:expo}. He uses the group for the
study of the exponential map associated to the pre-Lie operad. 
\end{Rm}
\begin{Tm}\label{Tm:charactergroup}
Let $P$ be a 1-reduced non-symmetric operad of finite type.
\begin{enumerate}
\item
The group 
$G_P$ is isomorphic to the group of characters of  the Hopf algebra $H_{P}$. 
\item
If $P$ is a 1-reduced operad of finite type, then the group
$\bar G_P$ is isomorphic to the group of characters of  the Hopf
algebra $\bar H_{P}$.  
\end{enumerate}
\end{Tm}
\begin{Pf}
The comultiplication on $H_P = T_*(\bigoplus_nP^*(n))$ induces a
multiplication on  
\[
\mathrm{Hom}_{\mathrm{Alg}}(H_P, k) \cong
\mathrm{Hom}_k(\bigoplus_nP^*(n), k)\cong \hat\bigoplus_{n\geq 2}P(n). 
\]
The $n\geq 2$ comes from the unitality of algebra homomorphisms:
the group consists of elements of $\hat\bigoplus_{n}P(n)$ such that
the coefficient for $\mathrm{id}^*$ equals 1.

Choose a homogeneous basis $\{p\}$ of $\bigoplus_nP(n)$ and let $\{p^*\}$ be
the dual basis. For such  basis elements with $p$ of degree $n-1$ and
$q$ and $r$ in arbitrary degree we have  
\begin{equation}\label{eq:dualcomform}
<r^*|p\cdot q> = <\gamma^*(r^*)|(p \otimes q^{\otimes n})> = 
<r^*|\gamma(p\otimes
q^{\otimes n})>
\end{equation}
with respect to the standard pairing $<.|.>$ between $P^*$ and
$P$. This shows the first part of the result. 
To obtain the statement on coinvariants, note that
\[
\Hom_{\mathrm{Alg}}(\bar H_P,k) = \hat\bigoplus_{n\geq 2}P(n)_{S_n},
\]
and replace $<.|.>$ in Equation (\ref{eq:dualcomform}) by the
pairing of $P(n)_{S_n}$ and $(P(n)^*)^{S_n}$. 
\end{Pf}
\begin{Rm}\label{Rm:symmofH}
Let $C$ be a cooperad. We want to be a bit more specific on the
relation between $H_C$ and $\bar H_C$ and their groups of
characters. Consider the symmetrisation $S_*(\bigoplus_nC(n))$ of
$H_C$. The formulae for the structure on $H_C$ 
make $S_*(\bigoplus_nC(n))$ a Hopf algebra.
Moreover, we have maps
\begin{equation}
H_C \longrightarrow S_*({\bigoplus}_nC(n)) \longleftarrow \bar H_C, 
\label{Rm:HandHbarmap}
\end{equation}
where the left map is a surjection and the right map is an
injection. A similar diagram exists for $B_C$ and $\bar B_C$.
Since the algebra $k$ is commutative, every character of $H_P$
factorises through its symmetrisation $S_*(\bigoplus_nP^*(n))$.
The quotient map $G_P \longrightarrow \bar G_P$ can thus be
interpreted as the map of character groups
\[
G_P  = \Hom_{\mathrm{Alg}}(S_*(\bigoplus_nP^*(n)), k) \longrightarrow
\Hom_{\mathrm{Alg}}(\bar H_P, k) = \bar G_P,
\]
induced by the map of Hopf algebras $\bar H_P \longrightarrow
S(\bigoplus_nP^*(n))$. 
\end{Rm}

\subsection{Lie algebras}

Let $P$ be a 1-reduced non-symmetric pseudo operad. The vector space $
L_P = \bigoplus_{n\geq 2}P(n)$ is a Lie algebra with respect to the
Lie bracket on $p\in P(n)$ and $q\in P(m)$ given by
\begin{equation}\label{Df:KapManLie}
[p,q] := \sum_{i=1}^n p\circ_i q - \sum_{j=1}^m q\circ_j p,
\end{equation}
where $p\circ_i q := p(\id^{\otimes i-1},q,\id^{\otimes n-i})$.
If $P$ is a 1-reduced pseudo operad, this Lie algebra structure
descends to the quotient $\bar L_P = \bigoplus_{n} P(n)_{S_n}$
(cf. Kapranov-Manin \cite{KaMa:ModMor}).  
Both Lie algebras are graded with respect to the grading 
$\mathrm{deg}(P(n)) = n-1$), and the application $P\longmapsto L_P$
(resp. $P\longmapsto \bar L_P$) defines a functor from non-symmetric
operads (resp. operads) to Lie algebras. 

\begin{Tm}\label{Tm:integrate}
Let $P$ be a 1-reduced non-symmetric operad of finite type. 
\begin{enumerate}
\item
The Lie algebra of primitive elements of $(H_P)^*$ is the Lie algebra 
$L_{P}$. 
\item
If $P$ is a 1-reduced  operad of finite type, the Lie algebra of
primitive elements of $(\bar H_P)^*$ is the Lie algebra $\bar L_{P}$. 
Consequently, $(\bar H_P)^*$ is the universal enveloping algebra
$U(\bar L_P)$ of the Lie algebra $\bar L_P$. 
\end{enumerate}
\end{Tm}
\begin{Pf}
Let $P$ be a 1-reduced non-symmetric operad of finite type. The Hopf
algebra $H_P^*$ is the pointed `cofree' coalgebra on the total space of
$P$ with the multiplication defined on $(P(k))\otimes 
(P(m_1)\otimes \ldots\otimes P(m_k))$ as the composition $\gamma$ of
the non-symmetric operad, and then extended as a coalgebra
homomorphism. Projected to the cogenerators $\bigoplus_nP(n)$, the
multiplication reduces to the sum of the circle-$i$ operations:
\[
\sum_{i=1}^m\circ_i : P(m)\otimes P(n) \longrightarrow P(m+n-1), 
\]
for $m,n>1$. The Lie bracket on the primitive elements is the
commutator of this (non-associative) product. This shows the first part
of the result. 

Assume that $P$ is a 1-reduced operad of finite type. The Lie algebra of
primitive elements of $(\bar H_P)^*$ is then the symmetric quotient
$\bar L_P$ since we have the factorisation through
$S_*(\bigoplus_nP(n))$ (cf. Remark \ref{Rm:symmofH}).  From the
Milnor-Moore theorem \cite{MilnorMoore:Hoa} it follows $H^*_P = U(\bar 
L_P)$.
\end{Pf}


\section{First examples}
\label{Sec:firstexamples}

\subsection{Formal diffeomorphisms}

\begin{Ex}
Consider the operad $\mathrm{Com}$, which has as algebras commutative
associative algebras. This operad satisfies $\mathrm{Com}(n) =
k$ for $n\geq 1$. Composition is the usual identification of tensor
powers of $k$ with $k$ itself. 

We now describe the Hopf algebra $H_{\mathrm{Com}}$.
Denote the generator of $\mathrm{Com}(n)$ by $e_{n}$. Thus
$H_{\mathrm{Com}}$ is the
pointed free associative algebra on generators $\{e_n\}_{n\geq 1}$
where $e_1$ is identified with the unit, and where $e_n$ is of degree
$n-1$, with coproduct 
\[
\Delta(e_{n}) = \sum_{k=1}^{n}\sum_{n_1+\ldots+n_k = n} e_{k}\otimes
e_{n_1}\cdot\ldots\cdot e_{n_k},
\]
where we sum over $n_i$ such that the formula makes
sense (i.e. $n_i\geq 1$). 
\end{Ex}

Let $\mathcal{H}^{\mathrm{diff}}$ be the pointed free commutative
algebra on variables $a_i$ for $i\geq 0$. The base-point is $1\longmapsto
a_0$. Define a bilinear pairing between the 
space of generators and the group of formal power series $\phi(x)$
with coefficients in $k$ such that
$\phi(x) \equiv x (\mathrm{ mod}\ x^2)$ by the formula 
\[
<a_n,\phi> = \frac{1}{(n+1)!}\frac{d^{n+1}}{dx^{n+1}}\phi(0).
\]
 Define a coproduct $\Delta:
\mathcal{H}^{\mathrm{diff}}\longrightarrow
\mathcal{H}^{\mathrm{diff}}\otimes  \mathcal{H}^{\mathrm{diff}}$ on
generators by the formula $<\Delta(a_n),f\otimes g> = <a_n ,f\circ
g>$. These formulae define a Hopf algebra. This Hopf algebra is the
\note{Hopf algebra of higher order differentials of the line}. The 
character group of $\mathcal{H}^{\mathrm{diff}}$ is the group of power
series $\phi(x)$ with coefficients in $k$ such that $\phi(x) \equiv x$
(mod $x^2$), which is also called the \note{group of formal diffeomorphisms
of the line}. The group multiplication is composition of power series.  

Define the Hopf algebra $\mathcal{H}^{\mathrm{ncdiff}}$ of
\note{non-commutative higher order differentials of the line}
(eg. Brouder-Frabetti \cite{BrouFrab:Noncom}). As an algebra
$\mathcal{H}^{\mathrm{ncdiff}}$ the free unital algebra on variables
$a_i$ for $i\geq 0$, with respect to the base-point $1 \longmapsto
a_0$. Define the coproduct on $\mathcal{H}^{\mathrm{ncdiff}}$ by
\[
\Delta(a_n) = \sum_{k,n-k=n_1+\ldots+ n_{k+1}} a_k \otimes
a_{n_1}\cdot\ldots\cdot a_{n_{k+1}},  
\]
where $k, n_i \geq 0$ for $i\leq k+1$.  (We got rid
of the usual binomial coefficients by introducing $1=a_0$ in the sum.)
Note that the symmetric quotient of $\mathcal{H}^{\mathrm{ncdiff}}$ is
$\mathcal{H}^{\mathrm{diff}}$.

\begin{Tm}
The completion (w.r.t. the $(\mathrm{arity} - 1)$-grading) of the Hopf
algebra $\bar H_{\mathrm{Com}}$ is the Hopf
algebra of higher order differentials of the line. Its character group
is $\bar G_{\mathrm{Com}}$, the group of formal diffeomorphisms of the
line. The Lie algebra of primitive elements of the dual $\bar
L_{\mathrm{Com}}$ is the Lie algebra of polynomial vector fields
without a constant term  on the line.   
\end{Tm}
\begin{Pf}
To identify the group $\bar G_{\mathrm{Com}}$ with the
group of power series in one variable $x$ with coefficients in $k$
such that  $\phi(x) \equiv x (\text{ mod} x^2)$ and use composition of
power series as 
multiplication. Use the isomorphism given by $e_n\longmapsto
x^n$. The result on the Hopf algebras now follows since a graded
complete commutative Hopf algebra of finite type is completely
determined by its group of characters (Quillen \cite{Qui:Rathtpy}). 

At the Hopf algebra level the isomorphism $\bar
H_{\mathrm{Com}}\longrightarrow \mathcal{H}^{\mathrm{diff}}$ is given on 
the basis of generators  by $e_n^*(\phi) \longmapsto a_{n-1}$. 

The Lie algebra $\bar L_{\mathrm{Com}}$ with basis $\{e_n\}_{n\geq 1}$
satisfies the commutation relation $[e_n,e_m] = (n-m)
e_{m+n-1}$. The explicit isomorphism is thus given by $e_n\longmapsto
x^n\pd_x$. 
\end{Pf}
It is clear that the map on the Hopf algebra level lifts to the
non-commutative version, which gives the following Corollory.
\begin{Cr}
The map defined on generators by $a_{i} \longmapsto e_{i+1}^*$ is an
isomorphism of graded Hopf algebras from $\mathcal{H}^{\mathrm{ncdiff}}$ to 
$H_{\mathrm{Com}}$.
\end{Cr}
\begin{Rm}
One can consider $\mathrm{Com}$ as the endomorphism operad of the one
dimensional vector space $k$. Analogous if one considers the
1-reduced version of the endomorphism operad $\End_V$ of a finite
dimensional vector space $V$, one obtains the higher dimensional analogue
of the higher order differentials, formal diffeomorphisms, and
polynomial vector fields. (Compare Kapranov-Manin
\cite{KaMa:ModMor}, where the result is stated on the level of Lie
algebras.)
\end{Rm}

\subsection{Associative and Lie operad}

\begin{Ex}
To describe the structures we get from the operad $\mathrm{Ass}$ of
associative algebra, first observe that the surjection $\mathrm{Ass}
\longrightarrow \mathrm{Com}$ becomes an isomorphism on coinvariants:
$\mathrm{Ass}(n)_{S_n} = \mathrm{Com}(n)_{S_n}$. thus $\bar H_{\mathrm{Ass}}
= \bar H_{\mathrm{Com}}$, $\bar L_{\mathrm{Ass}} = \bar L_{\mathrm{Com}}$,
and $\bar G_{\mathrm{Ass}} = \bar G_{\mathrm{Com}}$.   
However, in the non-symmetrised version there are some differences.

For the group $G_{\mathrm{Ass}}$ we write $x^{\sigma}$ for the element
corresponding to $\sigma\in S_n$.
The group $G_{\mathrm{Ass}}$ is then the group of formal
permutation-expanded series $\sum_{\sigma} c_\sigma x^{\sigma}$, where
$\sigma$ runs over permutations in $S_n$ for all $n$ and the
coefficients $c_\sigma$ are in the ground field. Moreover, the trivial
permutation $(1) \in S_1$ has coefficient $c_{(1)} = 1$. Composition
is the linear extension in $x^\sigma$ of
\[
x^\sigma \circ \left(\sum_\tau c_\tau x^\tau\right)
= \sum_{\tau_1,\ldots,\tau_k}
c_{\tau_1}\ldots c_{\tau_k} x^{\hat\sigma(\tau_1\times\ldots\times\tau_k)},
\]
where $\hat\sigma$ is the permutation that permutes $k$ blocks on
which the $\tau_i$ act. (Observe that $\hat\sigma$ thus depends on the
degree of the $\tau_i$.) The Lie algebra structure is given on
$\sigma\in S_n$ and $\tau\in S_m$ by
\[
[\sigma,\tau]  = \sum_{i=1}^n \sigma\circ_i \tau - \sum_{j=1}^m \tau \circ_j \sigma.
\]
Dually, the Hopf algebra $H_{\mathrm{Ass}} = T_*(\bigoplus_{n\geq 1}
S_n)$ with $S_n$ in degree $n-1$ has the coproduct
\[
\Delta(\sigma) = \sum_k\sum_{\sigma = \hat\tau_0\circ(\tau_1\times \ldots
\times \tau_k)}\tau_0 \otimes (\tau_1,\ldots,\tau_k),
\]
where the sum is over all decompositions of $\sigma$ as 
$\hat\tau_0 \circ (\tau_1,\ldots,\tau_k)$, where $(1)\in S_1$ is
identified with the unit in $H_{\mathrm{Ass}}$.
\end{Ex}
\begin{Lm}
Write $\mathbf{a}_1,\ldots,\mathbf{a}_n$ for the inputs. Then a basis
of $\mathrm{Lie}(n)$ is given by  
\begin{equation}\label{Eq:spanningset} 
\{ [\mathbf{a}_{\sigma(1)},[\mathbf{a}_{\sigma(2)},\ldots ,
[\mathbf{a}_{\sigma(n-1)},\mathbf{a}_n]...]]\qquad | \qquad \sigma \in
S_{n-1}\}, 
\end{equation}
where the brackets are in right-most position. 
\end{Lm}
\begin{Pf}
By a dimension argument
($\mathrm{dim}(\mathrm{Lie}(n)) = (n-1)!$) it suffices to show that
these elements span $\mathrm{Lie}(n)$. This can be proved by induction
on $n$, using the Jacobi identity.
\end{Pf}
%
%
\begin{Ex}
There is a natural inclusion of operads $\mathrm{Lie} \subset \mathrm{Ass}$
which is defined by sending the bracket $\lambda \in \mathrm{Lie}(2)$ to the 
commutator $\mu - \mu^{\mathrm{op}}$ of the associative product $\mu \in
\mathrm{Ass}(2)$.  The group $G_{\mathrm{Lie}}$ is therefore a subgroup of
$G_\mathrm{Ass}$. To make the group more explicit it is useful to
characterise the image of $\mathrm{Lie}$ in $\mathrm{Ass}$. Write
$\mathbf{a}_1,\ldots,\mathbf{a}_n$ for the inputs, then we can describe the
image of
$[\mathbf{a}_1,[\mathbf{a}_2,\ldots,[\mathbf{a}_{n-1},\mathbf{a}_n]...]]$
as the sum  
\[
\sum_{\sigma\in Z_n} (-1)^{n-\sigma^{-1}(n)}
\mathbf{a}_{\sigma(1)}\cdot \mathbf{a}_{\sigma(2)}\cdot\ldots\cdot
\mathbf{a}_{\sigma(n)},  
\]
where $Z_n$ consists of those permutations $\sigma\in S_n$ such 
that for $i := \sigma^{-1}(n)$,
\[
\sigma(1)<\sigma(2)\ldots<\sigma(i)\qquad \text{and}\qquad
\sigma(i)>\sigma(i+1)>\ldots>\sigma(n).
\]
By the Lemma above, the images of elements in
$G_{\mathrm{Lie}}$ are the series in $G_{\text{Ass}}$ that are of the form 
\[
\sum_n\sum_{\tau\in S_{n-1}} c_{n,\tau} \sum_{\sigma\in Z_n}
(-1)^{n-\sigma^{-1}(n)} x^{\sigma\circ (\tau\times (1))}.
\]
\end{Ex}

\subsection{The Connes-Kreimer Hopf algebra of trees}
\label{Df:ConnesKreimerhoaoftrees}
Consider the set $T(*)$ of rooted trees without external edges
different from the root. A \note{cut}
of $t\in T(*)$ is a subset of edges. A cut is \note{admissible} if
for every vertex $v\in\vert(t)$ the path from $v$ to the root contains
at most one edge in the cut.  If $c$ is an admissible cut of $t$,
denote by $R^c(t)$ the connected component of the graph obtained from
$t$ by removing the edges in $c$, where all new external edges are
removed. Denote by $P^c$ the disjoint union of the other
components as elements of $T(*)$ with their new root. 
\begin{figure}[!ht]
\centering
\input{treecut.pstex_t}
\caption{An admissible cut (on the left) and a non-admissible cut (on
the right). Construct $R^c(t)$ as the connected component containing the
root with all upward pointing external edges removed.}  
\end{figure}

The set $T(*)$ generates a commutative \hoa\ $\mathcal{H}_R$. As
an algebra it is the symmetric algebra on $T(*)$. Comultiplication
$\Delta$ is defined on generators $t$, and extended as a algebra
homomorphism: 
\begin{equation}\label{Eq:CKcoprod}
\Delta(t)=\sum_{c}R^c(t)\otimes P^c(t), 
\end{equation}
where the sum is over admissible cuts. 
The counit $\eps:\mathcal{H}_R\longrightarrow k$ takes value $1$ on
the empty tree and $0$ on other trees. The \hoa\ $\mathcal{H}_R$ is
the \note{Hopf algebra of rooted trees}, introduced by Kreimer
\cite{Krei:Hoa} and further studied in Connes-Kreimer
\cite{ConKr:Hoa}. 

The Lie algebra of primitive elements of $\mathcal{H}_R^*$ is
isomorphic to the  Lie algebra $\mathcal{L}_R$ spanned by rooted trees
without external edges other than the root, and the bracket $[s,t] =
s\bullet t - t\bullet s$, where
\[
t\bullet s = \sum_{v\in\vert(t)} t\circ_v s,
\]
and $t\circ_v s$ is the rooted tree consisting of the connected
subtrees $t$ and $s$ and one edge that connects the root of $s$ to the
vertex $v\in\vert(t)$.

\begin{Ex}
Let $M$ be the collection $M(n) = k$ for $n\geq 2$ and $M(n) = 0$
otherwise. Recall the free operad satisfies
$TM(n) = \bigoplus_{t\in T_2(\mathbf{n})} k$,
where $T_2(\mathbf{n})$ is the set of rooted trees with a 
a pointed bijection $\phi:\leg(t) \longrightarrow \{0,\ldots,n\}$
(i.e. the root leg is mapped to 0) where every vertex
has at least 3 external edges. 
Thus $TM$ is a graded operad of finite type with respect to
the grading by $|\vert(t)|$, the number of vertices. 
The symmetric group action permutes the labels on external edges.
\end{Ex}
\begin{Pp}
There exists an inclusion $\mathcal{H}_R \longrightarrow \widehat{\bar
H_{TM}}$, of the Connes-Kreimer Hopf algebra into the completion
(w.r.t. the grading by external edges) of $H_{TM}$. This inclusion is
graded with respect to the grading by the number of vertices.
\end{Pp}
\begin{Pf}     
It suffices to show that there is a surjection $\phi$ of Lie algebras $\bar
L_{TM}\longrightarrow \mathcal{L}_R$ onto the Lie algebra of primitive
elements of $\mathcal{H}^*_R$. 

A vertex $v\in
\vert(t)$ of a tree $t$ is called \note{saturated} if it has no external
edges other than the root attached to it. Denote by $J$ the ideal in
$TM$ spanned by trees with a saturated vertex. The map $\phi$ of Lie
algebras we define factorises through $\bar{L}_{TM/J}$. 
Thus is suffices to define $\phi: L_{TM}\longrightarrow \mathcal{L}_R$
on trees without a saturated vertex. On such a tree $t$ define 
\[
\phi(t) = \tilde t\cdot \prod_{v\in\vert(t)} i_t(v)!,
\]
where $i_t(v)$ is the number of external legs other than the root of
$t$ that are attached to vertex $v$ and $\tilde t$ is the tree $t$
with all external edges other than the root omitted. Recall that both
Lie algebras are pre-Lie algebras. To check that this is a Lie
algebra homomorphism, write
\[
\begin{split}
\phi(t)\bullet \phi(s) &= \sum_{u\in\vert(t)} \widetilde{t\circ_u s}  \cdot
\prod_{v\in\vert(t)}\prod_{w\in\vert(s)} i_t(v)! i_s(w)! \\
&= \sum_{u\in\vert(t)} \widetilde{t\circ_u s}  \cdot i_{t\circ_us}(u) \cdot
\prod_{v\in\vert(t\circ_u s)} i_{t\circ_u s}(v)!\\
&= \sum_{i=1}^{|\leg(t)|-1} \widetilde{t\circ_i s} \cdot
\prod_{v\in\vert(t\circ_u s)} i_{t\circ_u s}(v)!\\
&= \phi(t\bullet s).
\end{split}
\]
To understand the second equality, note that $i_s(w) = i_{t\circ_u
s}(w)$ for all $w\in \vert(s)$,  that $i_t(v) = i_{t\circ_u
s}(v)$ if $v\neq u$, and that $i_t(u) = i_{t\circ_u
s}(u)+1$. The third equality rewrites the sum over vertices as a sum
over external edges other than the root.
\end{Pf}

\subsection{The double symmetric algebra construction}

The results in this section first appeared in Van der Laan and
Moerdijk \cite{IekePep:Hopf}. We start with a lemma. This lemma is not
new, but we decided to include a sketch of the proof, since this point
has led to some confusion in earlier drafts. 
\begin{Lm}[Berger-Moerdijk \cite{BerMoer:Model}]
Let $A$ be a bialgebra. 
\begin{enumerate}
\item 
The vector spaces $C_A(n)= A^{\otimes n}$ (for $n\geq 1$) form a
coaugmented non-symmetric cooperad with as coidentity the counit
$\eps:A\longrightarrow k$ and the cocomposition $\gamma^*$ defined on
summands by the diagram
\[
\xymatrix{
A^{\otimes n} \ar@{.>}[r]^{\gamma^*\qquad}\ar[d]_{\Delta} & A^{\otimes k}
\otimes (A^{\otimes n_1}\otimes\ldots \otimes A^{\otimes n_k})\\
A^{\otimes n}\otimes A^{\otimes n} \ar@{=}[r] &(A^{\otimes
n_1}\otimes\ldots \otimes A^{\otimes n_k}) \otimes (A^{\otimes
n_1}\otimes\ldots \otimes A^{\otimes n_k}),
\ar[u]_{(\mu_1\otimes\ldots\otimes \mu_k)\otimes \id}
}
\]
where $\Delta$ is the coproduct of $A^{\otimes n}$, and
$\mu_i:A^{\otimes n_i}\longrightarrow A$ is the multiplication of the
algebra $A$. A coaugmentation for this non-symmetric cooperad is given
by the unit of $A$.
\item
The collection $C_A$ with the $S_n$-action by permuting tensor factors
is a coaugmented cooperad with respect to the same structure maps if
$A$ is commutative.
\end{enumerate} 
\end{Lm}
\begin{Pf}
Coassociativity follows directly from the fact that $\mu$ is a
coalgebra morphism. Coidentity and coaugmentation are also direct from
the bialgebra structure on $A$ and $A^{\otimes n}$.  Then it remains
to consider the compatibility with the $S_k\rtimes(
\bigoplus_{n_1+\ldots n_k = n}(S_{n_1}\times\ldots\times S_{n_k}))$-action for 
\[
\gamma^*:C_A(n) \longrightarrow \bigoplus_{n_1+\ldots + n_k = n} C_A(k)
\otimes C_A(n_1) \otimes \ldots \otimes C_A(n_k).
\]
In Sweedler's notation one can write the cocomposition $\gamma^*$ of $C_A$
on a generator $(x_1,\ldots,x_n)\in C_A(n)$ as
\[
\begin{split}
\gamma^*(x_1,\ldots,x_n) =
\sum\sum(x_1'\star\cdots\star
  x_{n_1}',\ldots,x_{n-n_k+1}'\star\cdots\star x_{n}') \otimes &\\
  ((x_1'',\ldots,x_{n_1}'')\otimes
  \ldots\otimes(x_{n-n_k+1}''\ldots,x_n''))&,
\end{split}
\]
where the first sum is over all $k$ and all partitions $n = n_1+
\ldots +n _k$, and the second sum is the sum of the Sweedler notation,
and where $\star$ denotes the product of $A$.
We study the action on the right hand side of this formula.
The compatibility with the $S_k$-action is satisfied since the action
permutes both the tensor factors in $A^{\otimes k}$, and the factors $A^{\otimes
n_1}$ up to $A^{\otimes n_k}$. For compatibility with the
$S_{n_i}$-action, consider  
$x_{n_1+\ldots+n_{i-1}+1}'\star\cdots\star x_{n_1+\ldots+n_{i}}'\otimes
(x_{n_1+\ldots+n_{i-1}+1}'',\ldots,x_{n_1+\ldots+n_{i}}'')$. The
$S_{n_i}$-action only permutes elements in the right hand side. Thus,
to obtain equivariance we should have that 
$x_{n_1+\ldots+n_{i-1}+1}'\star\cdots\star x_{n_1+\ldots+n_{i}}'$ is
invariant under permutation of the indices. In other words, the
product needs to be commutative.
\end{Pf}
Recall the bialgebras $T(\bar T'(A))$ and $S(\bar S'(A))$ (Brouder-Schmitt
\cite{Brou:Double}), where $T$ (resp. $S$) is the unital free associative
(resp. commutative) algebra functor, and $\bar T'$ (resp. $\bar S'$)
is the non-unital free  associative (resp. commutative) coalgebra
functor. Comparing the equation for $\gamma^*$ in Sweedler's notation
with the Brouder-Schmitt formulae makes the following result a tautology.
\begin{Tm}[Van der Laan-Moerdijk \cite{IekePep:Hopf}]\label{Pp:Brouder}
Let $A$ be a bialgebra. The bialgebra $T(\bar T'(A))$
is isomorphic to the opposite bialgebra of $B_{C_A}$.
The bialgebra $S(\bar S'(A))$ is isomorphic to the opposite bialgebra
of $\bar B_{C_A}$.
\end{Tm}
Let $C$ be a coaugmented cooperad. Then the collection $C^{>1}$
defined by $C^{>1}(1)= k$ and $C^{>1}(n) = C(n)$ for $n>1$ is a
1-reduced cooperad with cocomposition induced by cocomposition in
$C$ through the surjection $C \longrightarrow C^{>1}$.
\begin{Cr}[Van der Laan-Moerdijk \cite{IekePep:Hopf}]
The Pinter Hopf algebra associated to $T(\bar T'(A))$ (cf. Brouder and Schmitt
\cite{Brou:Double} for terminology) is isomorphic to the opposite Hopf algebra of
$H_{C^{>1}_A}$, and the Pinter Hopf algebra associated to $S(\bar S'(A))$
is isomorphic to the opposite Hopf algebra of $\bar H_{C^{>1}_A}$.
\end{Cr}

\section{Operads of graphs}

\label{Sec:operadofgraphs}
The examples treated in the previous section treat Hopf algebras based
on well known operads. However, a given Hopf algebra can also lead to
a new operad structure. The operads of graphs to which this section is
devoted are examples of this phenomenon.
\subsection{The operad $\Gamma$} 

In this section a labelled graph will be assumed to be a graph $\eta$
without edges from a vertex to itself, together with
a numbering of the vertices (i.e. a bijection
$\vert(\eta)\longrightarrow \{1,\ldots,|\vert(\eta)|\}$). 
If $k\leq |\vert(\eta)|$ denote by $\leg_\eta(k)$ the set of legs of
the vertex numbered $k$ in the labelled graph $\eta$. 
The restriction to labelled graphs without self-loops is not necessary, at this
point, but it catalyses some of the arguments.  

Define $\mathrm{Graph}(n)$ as the groupoid of labelled graphs $\eta$ such that 
$|\vert(\eta)| = n$ with isomorphisms of labelled graphs (compatible with the
numbering of vertices) as maps. 
\begin{Df}
Let $\Gamma(1) =k$ and 
$\Gamma(n) =\mathrm{colim}(\mathrm{Graph}(n))$ for $n\geq 2$. We define the
structure of an operad on the collection
$\Gamma$ defined by the $\Gamma(n)$ with the $S_n$-action on the
numbers of the vertices.

Let $\eta$ and $\zeta$ be two labelled graphs and let $k\leq
|\vert(\eta)|$. Assume that $|\leg_\eta(k)| = |\leg(\zeta)|$. For each
bijection $b:\leg_\eta(k)\longrightarrow \leg(\zeta)$ define $\eta\circ_b\zeta$
as the labelled graph defined by replacing vertex $k$ in $\eta$ by the
labelled graph
$\zeta$, and connecting the legs of $\zeta$ to the remaining part of
$\eta$ according to bijection $b$. The linear ordering of the vertices is
obtained from the linear ordering on the vertices of $\eta$ upon
replacing vertex $k$ by the linear order on vertices of $\zeta$. 
For $\eta$, $\zeta$ and $k$ as above define the circle-$k$ operation
of the operad as 
\begin{equation}\label{circiofGamma}
\eta\circ_k\zeta = \left\{\begin{array}{ l l} 
\sum_{[b]}\eta\circ_b \zeta & \qquad\text{if }|\leg_\eta(k)| = |\leg(\zeta)|\\
0 & \qquad \text{otherwise,}\end{array}\right.
\end{equation}
where the sum is over equivalence classes of bijections
$b:\leg_\eta(k)\longrightarrow \leg(\zeta)$ with respect to
the equivalence relation $b\sim b'$ iff $\eta\circ_b\zeta \cong \eta
\circ_{b'} \zeta$.

\begin{figure}[!ht]
\centering
\input{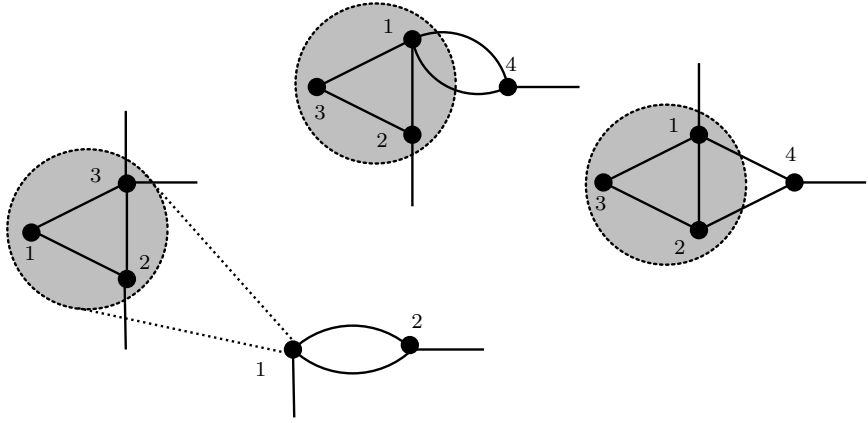}
\caption{The composition $\circ_1$ in the operad $\Gamma$. The graph
on the left is included in the two-vertex graphs. The result is the
sum of the two other graphs.}  
\end{figure}
\end{Df}
\begin{Lm}
The $\circ_k$-operations defined above, make the collection
$\Gamma$ an operad 
\end{Lm}
\begin{Pf}
Since the identities with respect to the $\circ_k$-operations are
trivial it remains to prove
associativity. Consider $(\eta\circ_k\zeta)\circ_l\theta$ for
$\eta,\zeta,\theta$ labelled graphs in $\Gamma$. If $l$ is a vertex of $\eta$
in $\eta\circ_k\zeta$, the labelled graphs $\zeta$ and $\theta$ are plugged
into different vertices of $\eta$. Thus we can write
\[
\begin{split}
(\eta\circ_k\zeta)\circ_l\theta &=
\sum_{[b],[b']}(\eta\circ_b\zeta)\circ_{b'}\theta\\
&= \sum_{[b],[b']}(\eta\circ_{b'}\theta)\circ_b \zeta,
\end{split}
\]
which is sufficient to prove associativity in this case.
Suppose on the contrary that $l$ is a vertex of $\zeta$ in
$\eta\circ_k\zeta$. Then 
\[
\begin{split}
(\eta\circ_k\zeta)\circ_l\theta &=
\sum_{[b]}(\eta\circ_b(\sum_{[b']}\zeta\circ_{b'}\theta))\\
&=\sum_{[b],[b']}\eta\circ_b(\zeta\circ_{b'}\theta).
\end{split}
\]
Regarding the second equality note that it is clear we can
move the sum over $[b']$ out since the automorphisms of labelled graphs have to
preserve the numbering of the vertices. 
\end{Pf}


\subsection{The Connes-Kreimer Hopf algebra of graphs}

Connes-Kreimer  \cite{ConKrei:RieHilI} introduces a Hopf algebra of
graphs (without numbered vertices) associated to the $\phi^3$ theory
in six dimensions. We define the Hopf algebra $\mathcal{H}_{\phi^3}$,
which is a slightly simplified version. 

A \note{1 particle irreducible graph} $\eta$  is a connected graph
$\eta$ such that $\eta$ has  at least two external edges, such that
each vertex $k\in\vert(\eta)$ satisfies $|\leg_\eta(k)|\in\{2,3\}$,
and such that the graphs are still connected after removing one
edge. This terminology comes from the physics. 
Let $\mathcal{H}_{\phi^3}$ as an algebra be the unital free
commutative algebra on the vector space spanned by 1 particle
irreducible graphs modulo the numbering of vertices $\bar\eta$. The
coproduct on a generator $\eta$ is given 
by
\[
\Delta(\bar\eta) = \sum_{\bar\zeta\subset\bar\eta} \bar\eta/\bar\zeta \otimes\bar \zeta,
\]
where the sum is over subgraphs $\bar\zeta$ such that $\bar\zeta$ is a direct
union of generators and if $v\in\vert(\bar\zeta)$, then
$\leg_{\bar\zeta}(v)= \leg_\eta(v)$. The direct union of generators
$\zeta$ in the right tensor factor is interpreted as the product of
generators. The counit is given by the usual augmentation of the
unital free commutative algebra.  

\begin{Df}
Let $\Gamma_{PI}$ be the suboperad of $\Gamma$ 
spanned by the 1 particle irreducible labelled graphs. It is easily checked
that this is indeed a suboperad. 
\end{Df}
\begin{Tm}\label{Tm:CKgraph}
The Hopf algebras $\bar H(\Gamma_{PI})$ and $\mathcal{H}_{\phi^3}$ are
isomorphic.
\end{Tm}
\begin{Pf}
By the Milnor-Moore Theorem, it suffices to compare the Lie algebras
of primitive elements of the dual Hopf algebras. In both cases the Lie
algebra of primitive elements is spanned by 1 particle irreducible
graphs without numbering of the vertices. 

If $\eta\in \Gamma_{PI}(n)$, denote by $\bar\eta\in
\Gamma(n)_{S_n}$ the graph $\eta$ obtained by forgetting the numbering
of vertices. In $P(\bar H(\Gamma_{PI})^*)$, the Lie bracket
(cf. Theorem \ref{Tm:integrate}) is given by
\[
[\bar\eta,\bar\zeta] = \sum_{v\in\vert(\bar\eta)}\sum_{[b]}
\bar\eta\circ_b \bar\zeta -
\sum_{w\in\vert(\bar\zeta)}\sum_{[b']} \bar\zeta\circ_{b'} \bar\eta \\
\]
where the sums in the first line are over equivalence classes of
bijections $b:\leg_{\bar\eta}(v)\longrightarrow \leg(\bar\zeta)$ for $v\in
\vert(\bar\eta)$ and 
$b:\leg_{\bar\zeta}(v)\longrightarrow \leg(\bar\eta)$ for $w\in
\vert(\bar\zeta)$.  
Theorem 2 in Connes-Kreimer \cite{ConKrei:RieHilI} states that this
Lie algebra is isomorphic to the Lie algebra of primitive elements in
$\mathcal{H}_{\phi^3}$. 
\end{Pf}
\begin{Rm}
A remark on the isomorphism is in place. 
Connes-Kreimer \cite{ConKrei:RieHilI} uses the notation
$\bar\eta\circ_v\bar\zeta := \sum_{v\in\vert(\bar\eta)}\sum_{[b]}
\bar\eta\circ_b \bar\zeta$, but do not state explicitly that they sum
over equivalence classes of bijections. The linear space
$P(\mathcal{H}_{\phi^3}^*)$ of primitive elements has a natural basis
spanned by graphs (dual to the basis of the generators of
$\mathcal{H}_{\phi^3}$ given above).

The Connes-Kreimer isomorphism
$P(\mathcal{H}_{\phi^3}^*)\longrightarrow \bar L_{\Gamma_{PI}}$
mentioned in the proof of Theorem \ref{Tm:CKgraph} is given by
\[
\bar\eta \longmapsto S(\bar\eta)\cdot\bar\eta,
\]
where $S(\bar\eta)$ is the symmetry factor of $\bar\eta$ which takes
the form
\[
S(\bar\eta) = \frac{|\Aut(\bar\eta)|}{|\Aut(\eta)|},
\]
where $\Aut(\eta)$ is the isomorphism group of $\eta$
(i.e. automorphisms need to preserve the numbering of vertices), and
$\Aut(\bar\eta)$ is the automorphism group of $\bar\eta$ (i.e. the
automorphism need not preserve the numbering of the vertices). In
other words, $S(\bar\eta)$ is the cardinality of the automorphism 
group of $\bar\eta$ divided by the subgroup of automorphisms that
induce the identity map on the vertices of $\bar \eta$.
Note that $S(\bar\eta)$ is indeed independent of the representative
$\eta$ of $\bar\eta$.  
\end{Rm}
\begin{Cr}
The group of characters of $\mathcal{H}_{\phi^3}$ is isomorphic to
$\bar G_{\Gamma_{PI}}$. It is the group of formal series of connected
1 particle irreducible graphs $\sum_{\bar\zeta} c_{\bar\zeta}
x^{\bar\zeta}$ with 
$c_\eta\in k$ and coefficient $c_{\bar\xi} = 1$ for each 1-vertex
graphs $\bar\xi$. The composition is the linear extension of 
\[
x^{\bar\eta}\ \cdot \ \sum_{\bar\zeta} d_{\bar\zeta}
x^{\bar\zeta} =
\sum_{\bar\zeta_1,\ldots,\bar\zeta_{|\vert(\bar\eta)|}}
d_{\zeta_1}\cdot\ldots\cdot d_{\zeta_{|\vert(\bar\eta)|}}
x^{\gamma(\bar\eta;\bar\zeta_1,\ldots,\bar\zeta_{|\vert(\bar\eta)|})},
\]
in terms of the composition of graphs in the operad $\Gamma_{PI}$,
where the one vertex trees are identified with the unit in the
algebra. 
\end{Cr}

\begin{Rm}
In order to reproduce the Hopf algebra originally defined by Connes
and Kreimer exactly, one needs to introduce a second colour for each
vertex with two legs, and one needs to label external edges by
elements of a certain space of distributions. 
\end{Rm}

\subsection{The operad $\tilde\Gamma$}

\begin{Df}
The definition of the operad $\Gamma$ above is not the only natural
choice. Let $\tilde\Gamma(n) = \Gamma(n)$ as an $S_n$-module for $n\in
\NN$. Define $\circ_k$-operations on $\tilde\Gamma$ as
\begin{equation}\label{circiofGammatilde}
\eta\circ_k\zeta = \left\{\begin{array}{ l l} 
\sum_b\eta\circ_b \zeta & \qquad\text{if }|\leg_\eta(k)| = |\leg(\zeta)|\\
0 & \qquad \text{otherwise,}\end{array}\right.,
\end{equation}
where $b$ runs over all bijections $b:\leg(\zeta)\longrightarrow
\leg_\eta(k)$, instead of equivalence classes (as in Formula
(\ref{circiofGamma})). 
\end{Df}
\begin{Pp}
The structure defined above makes $\tilde\Gamma$ an operad isomorphic
to $\Gamma$. 
\end{Pp}
\begin{Pf}
Checking the operad axioms for $\tilde\Gamma$ is not
difficult, the argument is analogous to the argument for
$\Gamma$. 

For a graph $\eta$, denote by $\Aut(\eta)$ the
automorphisms of $\eta$ that preserve the numbering of the vertices.
Define $\phi(\eta) = |Aut(\eta)|\cdot \eta $. This is an isomorphism of
collections from $\Gamma$ to $\tilde\Gamma$ (in characteristic 0). To
see that it commutes with the $\circ_i$-operations, one needs for
a bijection $b:\leg_\eta(i)\longrightarrow \leg(\zeta)$ the equality
\[
|\Aut(\eta\circ_b\zeta)|\cdot |\{b':\leg_\eta(i)\longrightarrow
\leg(\zeta)\ \mathrm{s.t.}\ [b']=[b]\}| = |\Aut(\eta)|\cdot |\Aut(\zeta)|.
\]
It suffices to construct a bijection
\[
\psi:\Aut(\eta)\times \Aut(\zeta)\longrightarrow \coprod_{b'\sim b}
\mathrm{Iso}(\eta\circ_b\zeta,\eta\circ_{b'}\zeta).
\]
Given $(\tau,\sigma)\in :\Aut(\eta)\times \Aut(\zeta)$, construct
$\psi(\tau,\sigma)$ as follows. There are natural inclusions of
$\zeta-\leg(\zeta)\hookrightarrow \eta\circ_{b'}\zeta$ and
$\eta-\leg_\eta(i) \hookrightarrow  \eta\circ_{b'}\zeta$ for every
$b'$. Let $b'=\sigma_{\leg(\zeta)}\circ b$, and define 
$\psi(\tau,\sigma): \eta\circ_b\zeta \longrightarrow
\eta\circ_{b'}\zeta$ by $\psi(\tau,\sigma)|_{\zeta-\leg(\zeta)} :=
\sigma|_{\zeta-\leg(\zeta)}$ and $\psi(\tau,\sigma)|_{\eta-\leg(i)} =
\tau|_{\eta-\leg(i)}$ on legs that are not glued. Denote the
leg obtained from glueing $l\in \leg_\eta(i)$ to $b(l)\in \leg(\zeta)$
by $\{l,b(l)\}$. On such legs define $\psi(\tau,\sigma)(\{l,b(l)\}) :=
\{\tau(l),b'(\tau(l)\}$. It remains to see that $\psi$ is a bijection.
 
If $\psi(\tau, \sigma) = \psi (\xi,\chi)$, then 
obviously $\tau|_{\eta-\leg(i)} = \xi|_{\eta-\leg(i)}$, and 
$\sigma|_{\zeta-\leg(\zeta)}= \chi|_{\zeta-\leg(\zeta)}$. Moreover,
$\sigma|_{\leg(\zeta)} = b'\circ b^{-1} = \chi|_{\leg(\zeta)}$, and
since $\{\tau(l), b'(\tau(l))\} = \{\xi(l),b'(\xi(l))\}$ it follows
that $\psi$ is an injection.

Given $\theta:\eta\circ_b\zeta \longrightarrow
\eta\circ_{b'} \zeta$, restriction to $\eta-\leg_{\eta}(i)$ and
$\zeta-\leg(\zeta)$ determine $(\tau,\sigma)\in
\Aut(\eta)\times\Aut(\zeta)$ except for glued legs. On these, let
$\sigma_{\leg(\zeta)} = b'\circ b^{-1}$, and if $\theta(\{l,b(l)\}) =
\{l',b'(l')\}$ define $\tau(l) = l'$. Then $\psi(\tau,\sigma) =\theta$
shows $\psi$ is surjective.
\end{Pf}

\subsection{The Wick algebra}

\begin{Df}
Let $V$ be a finite dimensional vector space together with a
quadratic form $b\in (V\otimes
V)^*$. Let $\eta$ be a graph with $n$ vertices. For $i
=1,\ldots,n$ denote $k_i = |\leg_\eta(i)|$, the number of legs of
vertex $i$, and write $S^{\prime \eta} V = \bigotimes_{i=1}^n
S^{\prime k_i} V$. For an edge $e\in \edge(\eta)$ with vertices
$j_1,j_2\in \vert(\eta)$ let $\eta - e$ denote the graph with edge $e$
omitted, and define
\[
b(e):S^{\prime \eta} \longrightarrow S^{\prime (\eta - e)}
\]
by applying $b$ to one tensor factor in $S^{\prime k_{j_1}}V$ and one tensor
factor in $S^{\prime k_{j_2}}V$. If $e'\neq e$ is an
other edge of $\eta$, then $e'\in\edge(\eta - e)$ implies that 
$b(e')\circ b(e):S^{\prime\eta} \longrightarrow S^{\prime(\eta - e
 -e')}$ is well defined and independent of the order of $e$ and
$e'$. 

Let $\eta$ be a graph with $n$ vertices and edges $\edge(\eta) =
\{e_1,\ldots,e_m\}$. For 
$i=1,\ldots,n$ denote by $k_i = |\leg_\eta(i)|$ the number of legs of
the vertex $i$, and by $l_i = |\edge_\eta(i)|$ the number of legs of
the vertex $i$ that are part of an edge. Define
$\tau^\eta:S^{\prime \eta} \longrightarrow S^{\prime |\leg(\eta|)}$ as 
\[
\tau^\eta = \sum_{\sigma\in S_{|\leg(\eta)|}}\sigma \circ
b(e_m)\circ\ldots\circ b(e_1).
\]
That is, first we contract tensor factors corresponding to all
edges, and on the result in $\bigotimes_{i=1}^nS^{\prime l_i}V$ we apply
the sum of all permutations to symmetrise the remaining tensor
factors, which means that we end up in $S^{\prime (l_1+\ldots+l_n)}V$. 
The result is again independent of the order of the edges. Extend
$\tau^\eta$ as zero to a map
\[
\tau^\eta:(S'V)^{\otimes n}\longrightarrow S'V.
\]
\end{Df}
\begin{Tm}\label{Pp:Wickcycalg}
Let $V$ be a vector space with a quadratic form $b\in (V\otimes V)^*$.
The symmetric coalgebra $S'V$ enjoys a $\Gamma$-algebra structure with
respect to the maps 
\[
\gamma(\eta,p_1,\ldots,p_{|\vert(\eta)|}) =
\frac{1}{|\Aut(\eta)|}\tau^\eta(p_1,\ldots,p_{|\vert(\eta)|}).
\]
\end{Tm}
\begin{Pf}
It suffices to show that 
$\tilde\gamma(\eta,p_1,\ldots,p_n) =\tau^{\eta}(p_1,\ldots,p_n)$
defines a $\tilde\Gamma$-algebra structure. Since compatibility with
the symmetric group action is obvious, it remains to check that
$\tau^{\eta}\circ_1\tau^{\zeta} =
\tau^{\eta\circ_1\zeta}$. 
Let $\edge(\eta) = \{e_1,\ldots,e_n\}$ and $\edge(\zeta) =
\{e_{n+1},\ldots,e_{n+m}\}$. For any
bijection $b:\leg(\zeta) \longrightarrow \leg_{\eta}(1)$ we can
naturally write $\eta\circ_b\zeta = \{e_1,\ldots,e_{n+m}\}$. 
Then 
\[
\tau^{\eta\circ_1\zeta} = \sum_{\sigma\in
S_{|\leg(\eta)|}} \sum_{b:\leg(\zeta) \longrightarrow \leg_{\eta}(1)} 
\sigma\circ b(e_{n+m})\circ\ldots\circ b(e_1),
\]
and
\[
\tau^{\eta}\circ_1\tau^{\zeta} = \sum_{\sigma\in
S_{|\leg(\eta)|}} \sum_{\tau\in S_{|\leg(\zeta)|}}
\sigma\circ b(e_{n})\circ \ldots b(e_1) \circ \tau \circ b(e_{n+m})
\circ \ldots \circ b(e_{n+1}).
\]
It is not hard to see that summing over $\tau$ and summing over $b$
have the same effect.
\end{Pf}

\begin{Rm}
Theorem \ref{Pp:Wickcycalg} is suggested by the Wick rotation
formula from quantum field theory (cf. Kazhdan \cite{Kaz:QFT}). In
particular by the asymptotic series for the functional integral of a
QFT in a neighbourhood of a free QFT.
\end{Rm}
\bibliographystyle{plain}
\bibliography{hopf}
~\\\textsc{Pepijn van der Laan} (\texttt{pvanderlaan@crm.es})\\ 
Centre de Recerca Matem\`atica, Apartat 50,  E-08139 Bellaterra 
\end{document}